\documentclass[journal, 10pt, twocolumn]{IEEEtran}

\usepackage{epsf}
\usepackage{graphics}
\usepackage{times}
\usepackage{graphicx}
\usepackage{amsmath}
\usepackage{amssymb}
\usepackage{epsfig}
\usepackage{subfigure}
\usepackage{array}
\usepackage{float}

\begin{document}

\title{PAPR Reduction in OFDM-IM Using Multilevel Dither Signals}

\author{Kee-Hoon Kim
\thanks{The author is with the Department of Electronic Engineering, Soonchunhyang University, Asan 31538, Korea (e-mail: keehk85@gmail.com).}}
\maketitle

\begin{abstract}
Orthogonal frequency division multiplexing with index modulation (OFDM-IM) is a novel multicarrier scheme, which uses the indices of the active subcarriers to transmit data. OFDM-IM also inherits the high peak-to-average power ratio (PAPR) problem, which induces in-band distortion and out-of-band radiation when OFDM-IM signal passes through high power amplifier (HPA). In this letter, dither signals are added in the idle subcarriers to reduce the PAPR. Unlike the previous work using single level dither signals, multilevel dither signals are used by exploiting that the amplitudes of the active subcarriers are variously distributed for different subblocks. As a result, the proposed scheme gives the dither signals more freedom (a larger radius of dithering) in average. Simulation results show that the proposed scheme can achieve better PAPR reduction performance than the previous work.
\end{abstract}


\section{Introduction}

Orthogonal frequency division multiplexing with index modulation (OFDM-IM) \cite{bacsar2013orthogonal} is a novel multicarrier technique, which extends the concept of spatial modulation (SM) \cite{mesleh2008spatial} into the frequency domain. In OFDM-IM, the subcarriers are divided into a series of subblocks, where the information bits are carried by two entities, the subcarrier indices and the modulated symbols. That is, the subcarriers have two states, active and idle, and the indices of the active subcarriers carry information.
The special design of OFDM-IM makes it more capable of combating inter-carrier interference (ICI) and gives better bit error rate (BER) performance in the low-to-medium data rate than the conventional OFDM \cite{bacsar2013orthogonal}. Also, it is possible to construct energy-efficient transmit signals compared to the conventional OFDM systems \cite{zhao2012high}.

As pointed out in \cite{ishikawa2016subcarrier}, OFDM-IM inherits the high peak-to-average power ratio (PAPR) problem from OFDM. As well known, high PAPR can cause in-band distortion and out-of-band radiation when the OFDM-IM signal passes through high power amplifier (HPA). Lots of PAPR reduction methods have been researched in OFDM literatures \cite{han2005overview}. To address the high PAPR problem in OFDM-IM, a way is to borrow the methods designed for OFDM directly. However, those methods may be not very efficient because they do not consider the unique characteristic of OFDM-IM structure \cite{zheng2017peak}.

Therefore, the authors in \cite{zheng2017peak} proposed an efficient PAPR reduction method through utilizing the idle subcarriers. Specifically, the scheme in \cite{zheng2017peak} introduces a dither signal with an amplitude constraint in the idle subcarrier, which is the first PAPR reduction method utilizing the characteristic of OFDM-IM.
However, this work does not consider the fact that the characteristics of the subblocks are various if quadrature amplitude modulation (QAM) is employed in the active subcarriers even though the index modulation and demodulation in OFDM-IM are performed subblock by subblock. It naively adopts the same amplitude constraint for every subblock.

In this letter, multilevel (multi amplitudes) dither signals are used, where the amplitude constraint can be varied and determined by the characteristic of each subblock. Specifically, the values of the modulated symbols in the subblock determine the according subblock's amplitude constraint. As a result, the proposed scheme gives the dither signals more freedom (a larger radius of dithering) in average and can provide better PAPR reduction performance than the previous work.

\section{System Model}
\subsection{Notations}
The superscript $T$ and $H$ denote the transpose and Hermitian transpose, respectively. $\mathbb{C}$ denotes the complex number domain. $|I|$ denotes the cardinality of the set $I$. Furthermore, $\|\zeta\|_{p}$ denotes the $p$-norm of the vector $\zeta$.
$\Re(\cdot)$ and $\Im(\cdot)$ present the real part and the imaginary part of the complex number, respectively.

\subsection{OFDM-IM}

In the OFDM-IM system, a total of $m$ information bits enter the OFDM-IM transmitter for the transmission of each OFDM-IM block. These $m$ bits are split into $g$ groups each containing $p$ bits, i.e., $m=pg$. Each group of $p$ bits is mapped to an OFDM subblock of length $n$ in frequency domain, where $n = N/g$ and $N$ is the number of subcarriers.
Unlike classical OFDM, this mapping operation is not only performed by means of the modulated symbols, but also by the indices of the subcarriers \cite{bacsar2013orthogonal}.

Inspired by the SM concept, additional information bits are transmitted by exploiting the OFDM subcarrier indices. For each
subblock, only $k$ out of $n$ available indices are activated for this purpose and they are determined by a selection procedure from a predefined set of active indices, based on the first $p_1$ bits of the incoming $p$-bit sequence.
We set the symbols corresponding to the inactive (or idle) subcarriers to zero, and therefore, we do not transmit data with them. The remaining $p_2=k \log_2M$ bits of this sequence are mapped onto the $M$-ary signal constellation to determine the symbols having active indices. Therefore, we have $p = p_1 + p_2$. In other words, in the OFDM-IM system, the information is conveyed by both of the $M$-ary modulated symbols and the indices of the modulated subcarriers (active subcarriers) \cite{bacsar2013orthogonal}.

Denote the set of the indices of the $k$ active subcarriers in the $\beta$-th subblock, $\beta=1,2,\cdots,g$, as
\begin{equation}
I_{\beta} = \{i_{\beta,1},i_{\beta,2},\cdots,i_{\beta,k}\}
\end{equation}
with $i_{\beta,\gamma} \in \{1,2,\cdots,n\}$ for $\beta = 1,2,\cdots,g$ and $\gamma = 1,2,\cdots,k$.
Correspondingly, the set of $k$ modulated symbols is denoted by
\begin{equation}
S_{\beta} =\{S_{\beta,1},S_{\beta,2},\cdots,S_{\beta,k}\}
\end{equation}
where $S_{\beta,\gamma} \in \mathcal{S}$ and $\mathcal{S}$ is the used signal constellation.

The OFDM-IM transmitter creates all of the subblocks by taking into account $I_{\beta}$ and $S_{\beta}$ for all $\beta$ first and it then forms the $N\times 1$ symbol sequence
\begin{equation}
\mathbf{X} = [X(1)~X(2)~\cdots~X(N)]^T
\end{equation}
where $X(\alpha) \in \{0, \mathcal{S}\},\alpha = 1,2,\cdots,N$, by concatenating these $g$ subblocks. Unlike the classical OFDM, in OFDM-IM $\mathbf{X}$ contains some zeros.
After these point, the same procedure as classical OFDM are applied. The symbol sequence in frequency domain is processed by the inverse discrete Fourier transform (IDFT) to generate the OFDM-IM signal sequence in time domain
\begin{equation}
\mathbf{x} = \mathbf{F}^H \mathbf{X}
\end{equation}
where $\mathbf{F}$ is the unitary DFT matrix and then $\mathbf{F}^H$ is the IDFT matrix. Then cyclic prefix (CP) is appended followed by parallel-to-serial (P/S) and digital-to-analog (D/A) conversion.
Also, the PAPR of the OFDM-IM signal sequence $\mathbf{x}$ is defined by
\begin{equation}
\mbox{PAPR}=\frac{\|\mathbf{x}\|_{\infty}^2}{\mathbb{E}[\|\mathbf{x}\|_2^2]/N}
\end{equation}
where $\mathbb{E}[\cdot]$ denotes the expectation operator.

The receiver's task is to detect the indices of the active subcarriers, which is called index demodulation in this letter, and demodulate the corresponding modulated symbols. There are several types of detection algorithms for the OFDM-IM system \cite{bacsar2013orthogonal}, \cite{siddiq2016low} and it is remarkable that the index demodulation is performed subblock by subblock.
After detection of the active indices by one of the detectors, the information is passed to the index demapper to provide an estimate of the $p_1$ bits. Once the active indices are determined, demodulation of the modulated symbols is straightforward and it provides an estimate of the $p_2$ bits.

\section{PAPR Reduction Using Multilevel Dither Signals}
\subsection{PAPR Reduction with Single Level Dither Signals \cite{zheng2017peak}}
First, we denote the set of the indices of the idle subcarriers in $\mathbf{X}$ as $U$, whose cardinality is $N-K$ and $K=kg$ is the number of active subcarriers in $\mathbf{X}$.
In \cite{zheng2017peak}, a dither signal with an amplitude constraint $R$ is introduced in the idle subcarrier and no signal change is required in the active subcarrier. The motivation of this amplitude constraint is  to control the impact of the dither signal on the index demodulation error. The design of the optimal dither signal can be solved by the convex programming.
That is, the PAPR optimization problem can be presented by
\begin{align}\label{eq:sdscvx}
\min_{\zeta} &\| \mathbf{F}_{U}^H \zeta + \mathbf{x} \|^2_{\infty}\nonumber\\
\mbox{s.t.}~&\|\zeta\|_{\infty} \leq R
\end{align}
where $\mathbf{x}=\mathbf{F}^H \mathbf{X}$ is the initial OFDM-IM signal sequence in time domain, $\mathbf{F}^H_{U}\in \mathbb{C}^{N\times (N-K)}$ is obtained from $\mathbf{F}^H$ by collecting the columns whose indices belong to $U$, and $\zeta$ is the $(N-K) \times 1$ dither signal in frequency domain. To solve the convex optimization problem, CVX \cite{grantcvx}, a package for specifying and solving convex programs, may be used.

\begin{figure}[htbp]
\centering
\includegraphics[width=\linewidth]{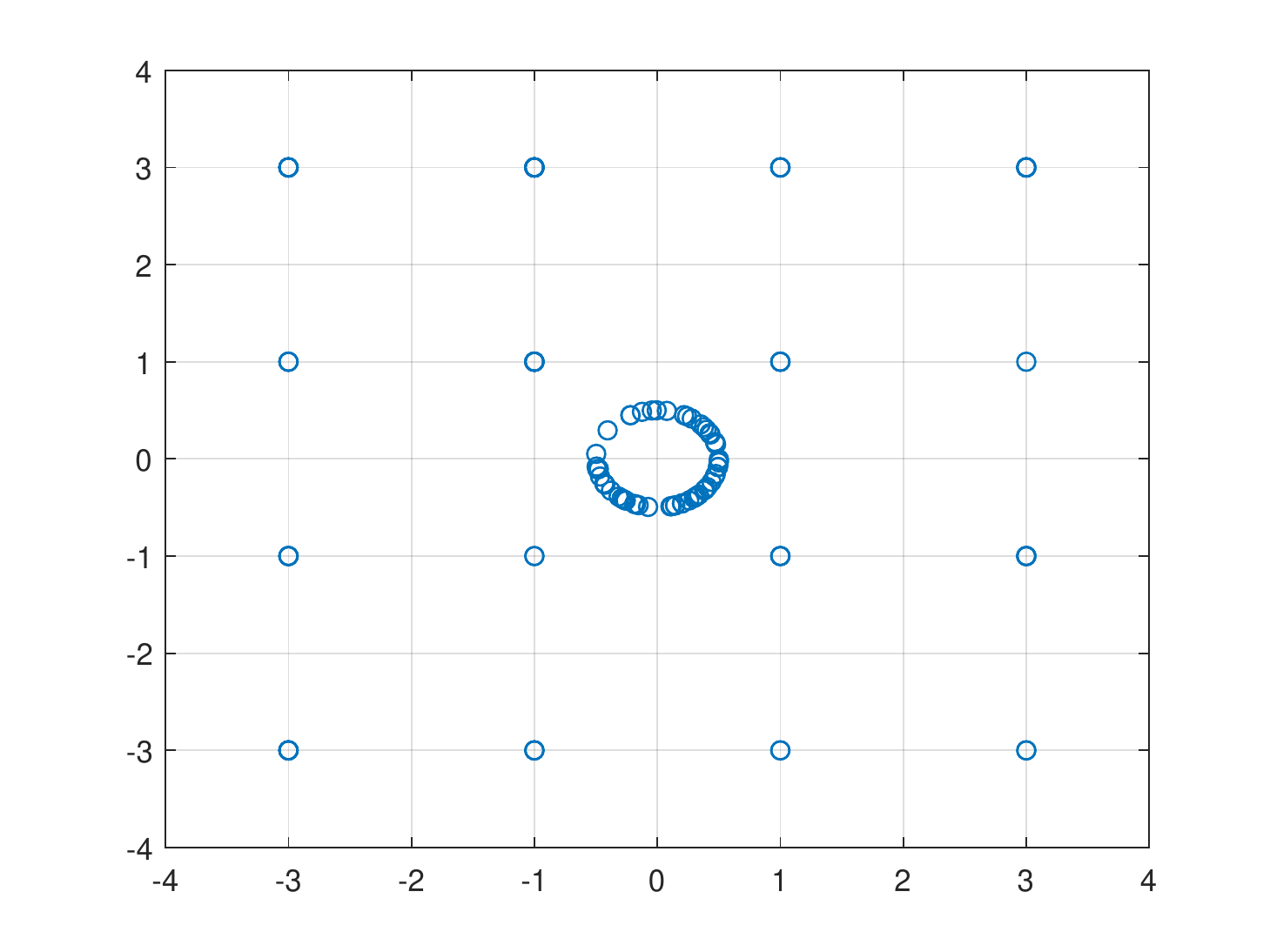}
\caption{The super constellation when 16-QAM is employed in the active subcarriers and the scheme in \cite{zheng2017peak} with $R=0.5$ is used. The original signal constellation points are $\Re(X(\alpha))\in\{\pm 1, \pm 3\}$ and $\Im(X(\alpha))\in\{\pm 1, \pm 3\}$.}
\label{fig:SDS}
\end{figure}

The super constellation after adding the dither signals is shown in Fig. \ref{fig:SDS}. Obviously, the signal in the active subcarrier keeps fixed, while the dither signals in the idle subcarrier are moved around the origin with the single amplitude constraint $R$.

\subsection{The Proposed PAPR Reduction with Multilevel Dither Signals}
\subsubsection{$\mu_{\beta}$, $\lambda_{\beta}$, $\delta_{\beta}$, and $\nu$}
In this subsubsection, we introduce four notations. First, the minimum amplitude of the modulated symbols in the active subcarriers in the $\beta$-th subblock is denoted by
\begin{equation}\label{eq:mubeta}
\mu_{\beta} = \min_{\gamma = 1,2,\cdots,k} |S_{\beta,\gamma}|
\end{equation}
where $|S_{\beta,\gamma}|$ denotes the amplitude of $S_{\beta,\gamma}$.
If we use QAM in the active subcarriers, the amplitude of the modulated symbols $|S_{\beta,\gamma}|$ is various. We denote the possible values of $|S_{\beta,\gamma}|$ as $A_1 < A_2 < \cdots < A_L$, where $L$ is the number of distinct amplitude levels. For example, let us consider 16-QAM, where $\Re(X(\alpha))\in\{\pm 1, \pm 3\}$ and $\Im(X(\alpha))\in\{\pm 1, \pm 3\}$. In this case, $L=3$ and we have $A_1 =\sqrt{1^2 + 1^2}=\sqrt{2}$, $A_2 =\sqrt{1^2 + 3^2}= \sqrt{10}$, and $A_3=\sqrt{3^2 + 3^2}=\sqrt{18}$. Therefore, $\mu_{\beta}$ in (\ref{eq:mubeta}) can also be varied for $\beta$ and be one of $\{A_1, A_2, \cdots, A_L\}$.

Second, the maximum amplitude of the dither signals in the idle subcarriers in the $\beta$-th subblock is denoted as $\lambda_{\beta}$. Without any dither signals, $\lambda_{\beta} = 0$ for all $g$ subblocks. Also, the scheme in \cite{zheng2017peak} guarantees $\lambda_{\beta} \leq R$ for all $g$ subblocks.

Third, the index demodulation error in the $\beta$-th OFDM-IM subblock mainly depends on the difference between $\mu_{\beta}$ and $\lambda_{\beta}$, which is denoted as
\begin{equation}
\delta_{\beta} = \mu_{\beta} - \lambda_{\beta}.
\end{equation}
As $\delta_{\beta}$ becomes smaller, the index demodulation error in the $\beta$-th subblock occurs more frequently.

If we take all $g$ subblocks into account, the minimum value of $\delta_{\beta}$ for all $g$ subblocks is dominant for index demodulation error in one OFDM-IM block, which is
\begin{equation}\label{eq:nu}
\nu=\min_{\beta =1,2,\cdots,g} \delta_{\beta}.
\end{equation}
That is, the index demodulation error occurs more frequently as $\nu$ in the OFDM-IM system becomes smaller. In conclusion, we have to keep $\nu$ as large as possible in OFDM-IM systems. Without any dither signals, it is clear that $\nu \geq A_1$ is guaranteed and the scheme in \cite{zheng2017peak} guarantees $\nu \geq A_1 - R$.

\subsubsection{The Proposed PAPR Reduction Scheme}
Using the fact that $\nu$ in (\ref{eq:nu}) is the minimum value among $g$ $\delta_{\beta}$'s and $\mu_{\beta}$ in $\delta_{\beta}$ is various for $\beta$, we propose the PAPR reduction scheme with multilevel dither signals. In the proposed scheme, the amplitude constraint in the idle subcarriers in the $\beta$-th subblock is determined according to $\mu_{\beta}$. Specifically, in the proposed scheme, the subblock whose $\mu_{\beta}$ is $A_l$ uses $R_l$, $l=1,2,\cdots,L$, as the amplitude constraint on the dither signals in the idle subcarriers.

To cast the convex problem, we introduce $L$ index sets $U_1, U_2, \cdots, U_{L}$ which are disjoint and the subsets of $U$. The $U_l$ consists of the indices of the idle subcarriers whose subblock has $\mu_{\beta} = A_l$.
Then, the PAPR optimization in the proposed scheme can be presented by
\begin{align}\label{eq:mdscvx}
\min_{\zeta_1, \zeta_2, \cdots, \zeta_{L}} &\| \mathbf{F}_{U_1}^H \zeta_1 + \mathbf{F}_{U_2}^H \zeta_2 + \cdots + \mathbf{F}_{U_{L}}^H \zeta_{L} + \mathbf{x} \|^2_{\infty}\nonumber\\
\mbox{s.t.}~&\|\zeta_1\|_{\infty} \leq R_1\nonumber\\
&\|\zeta_2\|_{\infty} \leq R_2\nonumber\\
&~~~\vdots\nonumber\\
&\|\zeta_{L}\|_{\infty} \leq R_{L}
\end{align}
where $\mathbf{F}_{U_l}^H\in \mathbb{C}^{N\times |U_l|}$ is obtained from $\mathbf{F}^H$ by collecting the columns whose indices belong to $U_l$ and $\zeta_l$ is the dither signals with length $|U_l|$, $l=1,2,\cdots,L$.
Note that the computational complexity to solve the convex problem in (\ref{eq:mdscvx}) is the same as that for (\ref{eq:sdscvx}) because the dimensions of the variables and the constraints are the same in total.

Now, in order to guarantee $\nu \geq c$ and give the dither signals more freedom at the same time, the following criterion may be the optimal one;
\begin{equation}\label{eq:cri}
A_1 - R_1 = A_2 - R_2 = \cdots = A_L - R_L = c.
\end{equation}
By using this criterion, in the subblock having large $\mu_{\beta}$, the dither signals with large amplitudes may be added in the idle subcarriers.

\subsubsection{Comparison between the Proposed Scheme and the Scheme in \cite{zheng2017peak}}

As we mentioned, the scheme in \cite{zheng2017peak} guarantees $\nu \geq A_1 - R$. If we set $R_1 = R$ in the proposed scheme, the proposed scheme also guarantees $\nu \geq A_1 - R$. However, it is clear that the proposed scheme gives the dither signals more freedom because $R_l = A_l - A_1 + R_1 > R_1$ for $l \neq 1$ from (\ref{eq:cri}).
In this case, the proposed scheme can achieve the better PAPR reduction performance while keeping the index demodulation error the same compared to the scheme in \cite{zheng2017peak}.

\begin{figure}[htbp]
\centering
\includegraphics[width=\linewidth]{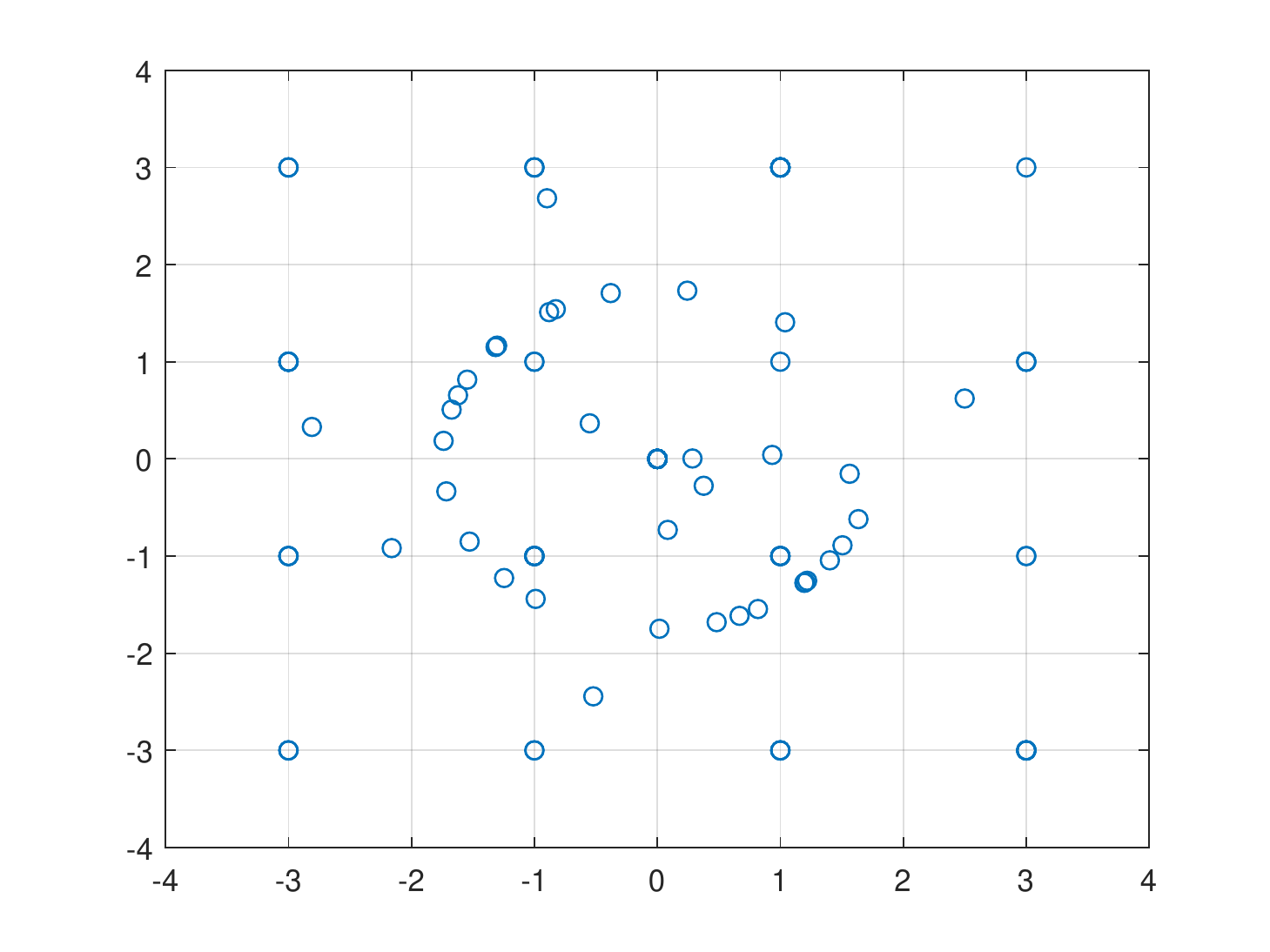}
\caption{The super constellation when 16-QAM is employed in the active subcarriers and the proposed scheme with $R_1=0$, $R_2=1.75$, and $R_3=2.83$ is used. The original signal constellation points are $\Re(X(\alpha))\in\{\pm 1, \pm 3\}$ and $\Im(X(\alpha))\in\{\pm 1, \pm 3\}$.}
\label{fig:MDS}
\end{figure}

The super constellation of the proposed scheme is shown in Fig. \ref{fig:MDS}. We obey the criterion in (\ref{eq:cri}) and use $R_1 = 0$, which means $R_2 = A_2 -A_1 = \sqrt{10}-\sqrt{2} \simeq 1.75$ and $R_3 = A_3 - A_1 = \sqrt{18}-\sqrt{2} \simeq 2.83$. (In simulations, we found that even $R_1=0$ is enough to provide good PAPR reduction performance.)
Obviously, the signal in the active subcarrier keeps fixed, while the signal in the idle subcarrier can be moved around the origin with various amplitude constraints $R_1, R_2, \cdots, R_L$. Also, it is clear that if the constant amplitude modulation such as quadrature phase shift keying (QPSK) is used, the proposed scheme is the same as the scheme in \cite{zheng2017peak}.

\section{Simulation Results}
In this section, we present our simulation results of the PAPR reduction and BER performance of the proposed scheme, the scheme in \cite{zheng2017peak}, and the original OFDM-IM signal. We employ the low complexity power based index demodulation algorithm \cite{siddiq2016low}, where the elements with $k$ highest powers in the subblock are considered as the active subcarriers. We use $N=128$, $n=4$, and $k=2$. For modulating the symbols in the active subcarriers, 16-QAM is used, where $\Re(X(\alpha))\in\{\pm 1, \pm 3\}$ and $\Im(X(\alpha))\in\{\pm 1, \pm 3\}$.

The scheme in \cite{zheng2017peak} uses $R=0.5$. The proposed scheme uses $R_1=0$ and the other $R_l$'s for $l\neq 1$ are selected to satisfy the criterion in (\ref{eq:cri}) because we found that $R_1=0$ in the proposed scheme is enough through several simulations. Even $R_1=0$ means the idle subcarriers whose indices are in $U_1$ cannot have any dither signals, instead $R_2$ and $R_3$ are large enough ($R_2= 1.75$ and $R_3= 2.83$).

\begin{figure}[htbp]
\centering
\includegraphics[width=\linewidth]{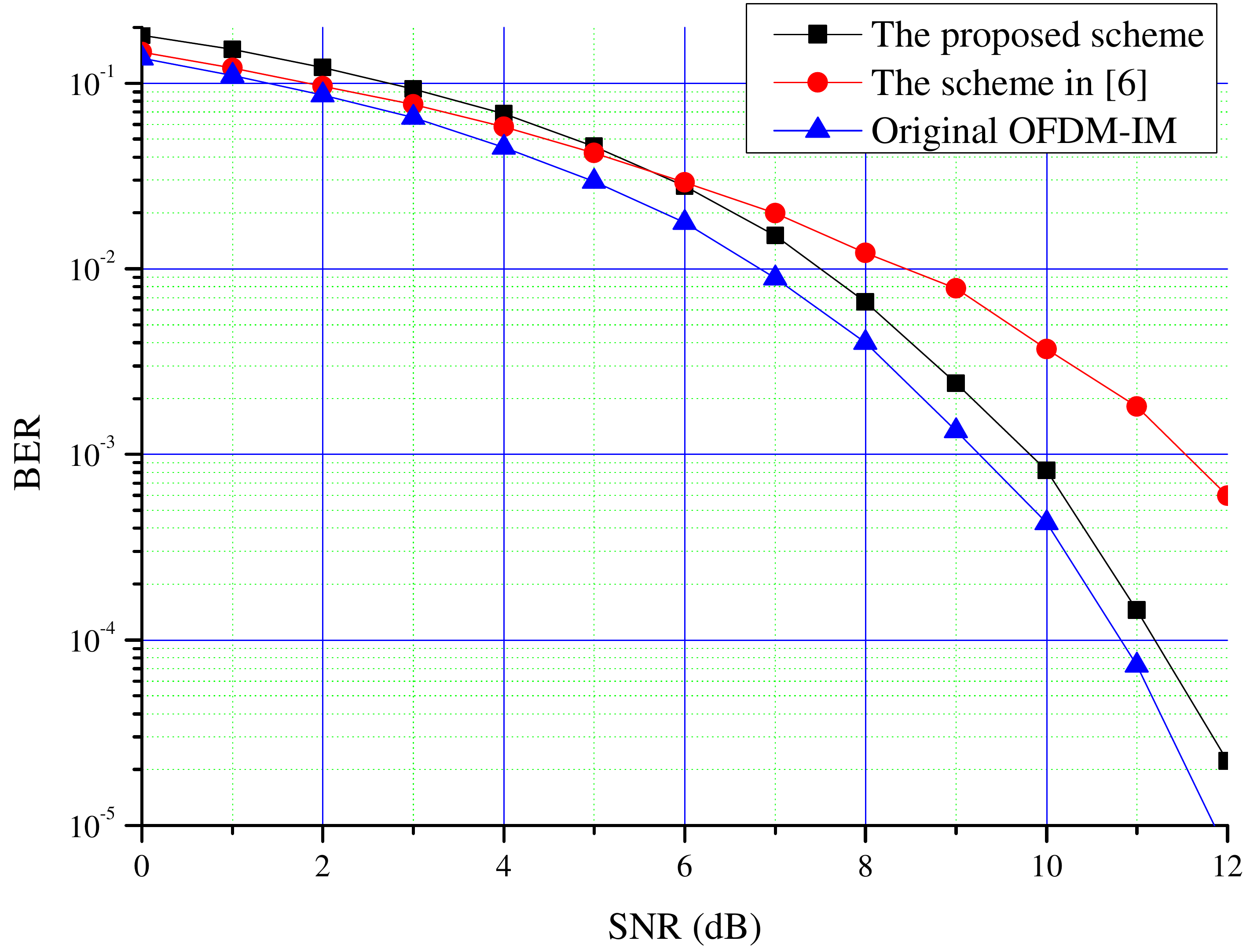}
\caption{The BER performance of the proposed scheme, the scheme in \cite{zheng2017peak}, and the original OFDM-IM signal.}
\label{fig:BER}
\end{figure}

Fig. \ref{fig:BER} shows the BER performance of the three schemes. The SNR means the average energy per bit over the additive white Gaussian noise power. In Fig. \ref{fig:BER}, the proposed scheme performs well even though there is a degradation about 0.6dB in comparison with the original OFDM-IM. This is mainly comes from the fact that the dither signals in the idle subcarriers in the proposed scheme increase the transmit power of the OFDM-IM signal and then the average energy per bit in SNR increases. (From simulations, the average energy per bit of the proposed scheme is 0.6dB larger than that of the original OFDM-IM signal.) Note that the proposed scheme has the same lower bound $\nu \geq A_1$ as the original OFDM-IM signal when the proposed scheme uses $R_1 = 0$. However, the scheme in \cite{zheng2017peak} shows a bad BER degradation especially for high SNR region because it has the smaller lower bound $\nu \geq A_1 - R = A_1 - 0.5$ than the other schemes.

\begin{figure}[htbp]
\centering
\includegraphics[width=\linewidth]{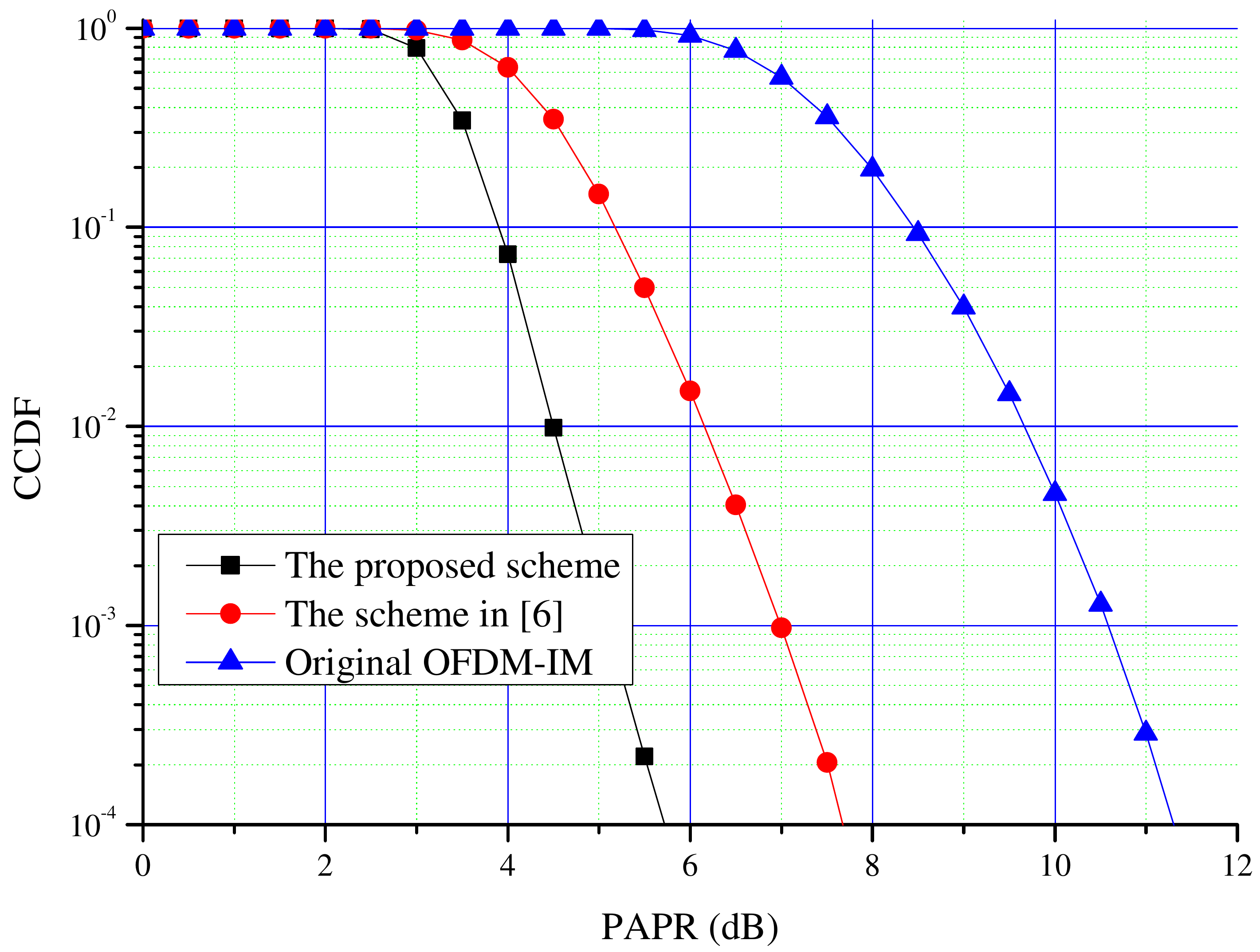}
\caption{The PAPR performance of the proposed scheme, the scheme in \cite{zheng2017peak}, and the original OFDM-IM signal.}
\label{fig:PAPR}
\end{figure}

Fig. \ref{fig:PAPR} shows the PAPR reduction performance of the three schemes. The ordinate is the complementary cumulative distribution function (CCDF) of the PAPR. In Fig. \ref{fig:PAPR}, the proposed scheme shows the best PAPR reduction performance because it has the largest freedom of dithering in the idle subcarriers in average. Even $R_1=0$ means the idle subcarriers whose indices are in $U_1$ cannot have any dither signals, instead $R_2$ and $R_3$ are large enough.

\section{Conclusion}
In this letter, the novel PAPR reduction scheme for OFDM-IM using the multilevel dither signals is proposed. By giving the variable amplitude constraints to the idle subcarriers according to the characteristic of each subblock, the freedom of dithering in the idle subcarriers is maximized while keeping $\nu$ as large as possible. As a result, the proposed scheme has much better PAPR reduction performance than the previous work with a slight BER performance degradation compared to the original OFDM-IM signal case.

\section*{Acknowledgment}
Acknowledgment

\bibliographystyle{IEEEtran}
\bibliography{biblio,IEEEfull}

\end{document}